\documentclass[%
 aip,
 amsmath,amssymb,
 reprint,%
]{revtex4-1}

\usepackage{graphicx}
\usepackage{dcolumn}
\usepackage{bm}

\usepackage[utf8]{inputenc}
\usepackage[T1]{fontenc}
\usepackage{mathptmx}
\usepackage{tabularx}
\usepackage{longtable, tabu}
\usepackage{booktabs}
\usepackage{multirow}
\usepackage{soul}
\usepackage{filecontents}

\usepackage[separate-uncertainty = true,
            multi-part-units=single,
            per-mode=symbol]{siunitx}
\DeclareSIUnit\sq{\ensuremath{\Box}}

\usepackage{mhchem}

\begin{document}

\preprint{AIP/123-QED}

\title[]{Magnetic field resilient high kinetic inductance superconducting niobium nitride coplanar waveguide resonators}

\author{Cécile Xinqing Yu}
 \thanks{Authors to whom correspondence should be addressed: cecile.yu@cea.fr, romain.maurand@cea.fr}
 \affiliation{Univ. Grenoble Alpes, CEA, Grenoble INP, IRIG, PHELIQS, F-38000 Grenoble, France}

\author{Simon Zihlmann}%
 \affiliation{Univ. Grenoble Alpes, CEA, Grenoble INP, IRIG, PHELIQS, F-38000 Grenoble, France}

\author{Gonzalo Troncoso Fernández-Bada}
 \affiliation{Univ. Grenoble Alpes, CEA, Grenoble INP, IRIG, PHELIQS, F-38000 Grenoble, France}
 
\author{Jean-Luc Thomassin}
 \affiliation{Univ. Grenoble Alpes, CEA, Grenoble INP, IRIG, PHELIQS, F-38000 Grenoble, France}
 
 \author{Frédéric Gustavo}
 \affiliation{Univ. Grenoble Alpes, CEA, Grenoble INP, IRIG, PHELIQS, F-38000 Grenoble, France}
 
\author{\'{E}tienne Dumur}
 \affiliation{Univ. Grenoble Alpes, CEA, Grenoble INP, IRIG, PHELIQS, F-38000 Grenoble, France}

\author{Romain Maurand}
 \thanks{Authors to whom correspondence should be addressed: cecile.yu@cea.fr, romain.maurand@cea.fr}
 \affiliation{Univ. Grenoble Alpes, CEA, Grenoble INP, IRIG, PHELIQS, F-38000 Grenoble, France}

\date{2 February 2021}

\begin{abstract}

We characterize niobium nitride (NbN) $\lambda/2$ coplanar waveguide resonators, which were fabricated from a \SI{10}{\nano\meter} thick film on silicon dioxide grown by sputter deposition. For films grown at \SI{180}{\celsius} we report a superconducting critical temperature of \SI{7.4}{\kelvin} associated with a normal square resistance of \SI{1}{\kilo\ohm} leading to a kinetic inductance of \SI{192}{\pico\henry\per\sq}. We fabricated resonators with a characteristic impedance up to \SI{4.1}{\kilo\ohm} and internal quality factors $Q_\mathrm{i} > 10^4$ in the single photon regime at zero magnetic field. Moreover, in the many photons regime, the resonators present high magnetic field resilience with  $Q_\mathrm{i} > 10^4$ in a \SI{6}{\tesla} in-plane magnetic field as well as in a \SI{300}{\milli\tesla} out-of-plane magnetic field. These findings make such resonators a compelling choice for cQED experiments involving quantum systems with small electric dipole moments operated in finite magnetic fields.
\end{abstract}

\maketitle

High quality superconducting microwave resonators are at the heart of circuit quantum electrodynamics (cQED) experiments \cite{Blais2004, Blais2020, Clerk2020}.
In recent years, high-impedance superconducting microwave resonators (HISMRs) have emerged as a new component\cite{Samkharadze2016, Grunhaupt2019, Niepce2019} allowing to explore regimes previously unattained \cite{Mi2018, Samkharadze2018, Landing2018,Grunhaupt2019}.
Such resonators are characterized by a characteristic impedance $Z_\mathrm{c}$ being close or even higher than the quantum of resistance $h/(2e)^2 \approx\SI{6.5}{\kilo\ohm}$.
To reach such a high impedance, the resonators need low stray capacitance and extremely high inductance\cite{Masluk2012, Bell2012} as $Z_\mathrm{c}=\sqrt{L_\ell/C_\ell}$ with $L_\ell$ ($C_\ell$) the inductance (capacitance) per unit length.
A large inductance can be achieved either with Josephson meta-materials \cite{B.D.Josephson1962, Castellanos-Beltran2007}, or thanks to the large kinetic inductance of disordered superconductors like TiN\cite{Sharrow2018}, NbTiN \cite{Samkharadze2016} or granular aluminium\cite{Grunhaupt2019}.
Through their high impedances, HISMRs generate large zero-point voltage fluctuations $V_\mathrm{ZPF} \propto f_0 \sqrt{Z_\mathrm{c}}$ with $f_0$ the resonator fundamental frequency.
A large $V_\mathrm{ZPF}$ is the key parameter in the coupling of microwave photons to small electrical dipole moments like polar molecules \cite{Andre2006} or charges in semiconductor quantum dots \cite{Delbecq2011, Frey2012}. Enhancing the coupling strength of such hybrid systems requires resonators with a large $V_\mathrm{ZPF}$ and hence a large $Z_\mathrm{c}$\cite{Stockklauser2017}.

Moreover, the magnetic field resilience of the HISMRs is essential for cQED type experiments with quantum systems requiring magnetic fields like spin \cite{Viennot2015,Mi2018,Borjans2019} or majorana fermion qubits \cite{Muller2013,Yavilberg2015,Dartiailh2017}. Different strategies are explored to maintain the quality factors of superconducting resonators in magnetic fields such as vortices traps in the milli-teslas range\cite{Bothner2011,deGraaf2012,Bothner2017, Kroll2019} or using disordered superconductors with a high critical magnetic field\cite{Samkharadze2016,Kroll2019,Zollitsch2019,Borisov2020} in the several teslas range.

In this prospect, we present superconducting microwave $\lambda/2$ coplanar waveguide (CPW) resonators made from \SI{10}{\nano\meter} thick films of NbN.
We first study the kinetic inductance of the films by four probe DC measurement and by two-tone spectroscopy on long resonators.
We show that the substrate temperature during the film growth is a viable control knob to achieve a desired kinetic inductance value.
We fabricate, in one etching step, resonators of different characteristic impedances ranging from \SI{110}{\ohm} to \SI{4.1}{k\ohm}, just by varying the geometry of the CPW.
We characterize the internal quality factor of the resonators as a function of the average photon number occupancy.
Then we extend the investigation by studying the resilience of the resonators to in-plane and out-of-plane magnetic fields.
Finally, we pinpoint that the \SI{4.1}{\kilo\ohm} resonators, which induce the highest zero-point voltage fluctuations, show $Q_\mathrm{i} > 10^4$ in the single photon regime, while preserving a high quality factor in both \SI{300}{\milli\tesla} out-of-plane and \SI{6}{\tesla} in-plane magnetic fields. 
This makes NbN HISMR a compelling choice for cQED experiments involving quantum systems with small electric dipole moments under sizable magnetic fields.

\begin{figure}[h!]
    \centering
    \includegraphics{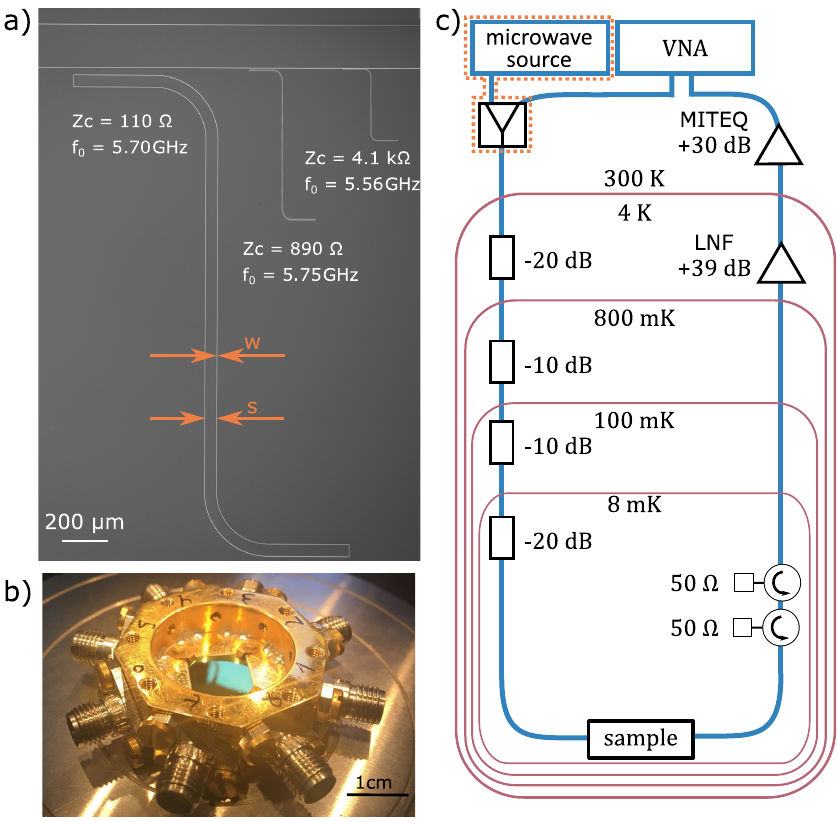}
    \caption{Experimental implementation.
    a) Three combined SEM images of the NbN CPW resonators with different gap and central conductor width, $w$ and $s$ respectively.
    b) Image of a chip bonded inside a sample holder.
    c) Experimental set-up schematic used to perform microwave transmission measurements. The part enclosed by orange dashed line is only used for two-tone spectroscopy.}
    \label{fig:setup}
\end{figure}

The resonators are fabricated on a \SI{525(25)}{\micro\meter}
thick p-type silicon wafer (1-\SI{15}{\ohm\centi\meter}), covered by \SI{400(80)}{\nano\meter} of thermally grown \ce{SiO2}.
The NbN deposition is performed using a Plassys MP600S confocal sputter system where the wafer is first heated during $\sim$\SI{16}{\hour} at \SI{180}{\celsius} at a base pressure of \SI{2e-9}{\milli\bar}.
Then, we perform a cleaning step of \SI{30}{s} of Ar milling with a bias voltage of \SI{350}{V}. The sputtering step lasts for \SI{11}{s} to deposit \SI{10}{\nano\meter} of NbN by DC magnetron sputtering at \SI{0.01}{\milli\bar} with an Ar:N partial gas ratio of [60:40].
Afterwards the resonators are patterned in one e-beam lithography step using the ZEP resist, followed by an O\textsubscript {2}/SF\textsubscript{6} plasma etching.

Representative scanning electron microscopy (SEM) images of the resonators obtained after such a process are shown in Fig.~\ref{fig:setup}(a).
We designed arrays of resonators in a hanger type geometry allowing parallel measurements of 5 resonators in one experimental run.
In Fig.~\ref{fig:setup}(b), we show an instance of such a chip inside its sample box.
The sample box is placed in a dry dilution refrigerator equipped with a 3D vector magnet (6 - 1 - \SI{1}{\tesla}) and connected to a standard microwave setup, see Fig.~\ref{fig:setup}(c).
The chip is then cooled down to a base temperature of \SI{8}{\milli\kelvin} at zero magnetic field.

\begin{figure}
\includegraphics{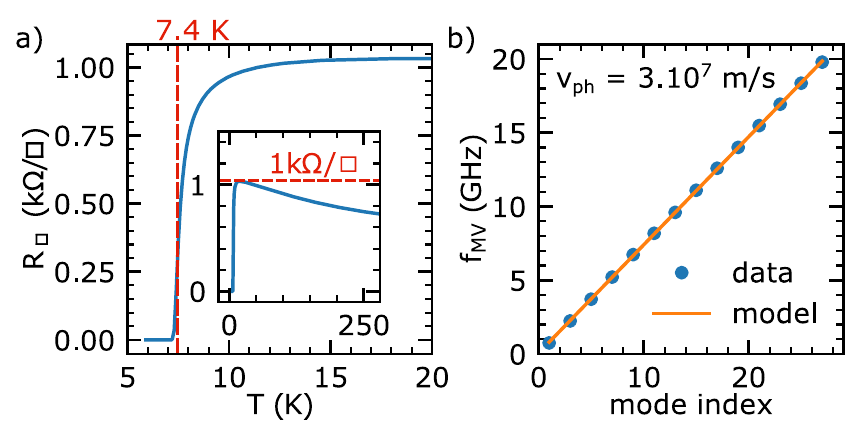}
\caption{\label{fig:kinetic} DC and RF method to extract the kinetic inductance.
a) R(T) characteristics of the NbN film, $T_c = \SI{7.40}{\kelvin}$ and $R_{\Box} = \SI{1033}{\ohm\per\sq}$.
b) Dispersion relation of a NbN resonator measured by two-tone spectroscopy, only half of the data points to extract the phase velocity is plotted.}
\end{figure}

We characterize the NbN film by measuring its sheet resistance as a function of temperature, see Fig.~\ref{fig:kinetic}(a).
The inset shows the high temperature behaviour for which the sheet resistance increases while lowering the temperature from room temperature to $\sim\SI{19}{\kelvin}$, typical of weak localization and Coulomb interaction in strongly disordered superconductors\cite{Sacepe}.
From 19 to \SI{5}{\kelvin} the resistance decreases, until zero resistance marking the superconducting transition.
From this curve we extract the sheet resistance $R_\mathrm{\Box} = \SI{1033(1)}{\ohm\per\sq}$ as the maximal value of the curve and the critical temperature $T_\mathrm{c} = \SI[parse-numbers = false]{7.4 \pm 0.1}{\kelvin}$ as the temperature at the inflexion point of the curve, see Fig.~\ref{fig:kinetic}(a).
From the sheet resistance and the critical temperature, we estimate the kinetic inductance of our NbN film\cite{Tinkham} as
\begin{equation}
L_\mathrm{kin} = \frac{\hbar R_\mathrm{\Box}}{\pi \Delta_0}, 
\end{equation}
where $\hbar$ is the reduced Planck constant and $\Delta_0$ is the superconducting gap at zero temperature.
We assume that the superconducting gap for NbN is given by $\Delta_0 = 1.76k_\mathrm{B}T_\mathrm{c}$ where $k_\mathrm{B}$ the Boltzmann constant \cite{Antonova1981}.
From this DC measurement we obtain a kinetic inductance value of $L_\mathrm{kin} = \SI{192(3)}{\pico\henry\per\sq}$.

To confirm the kinetic inductance value extracted via DC measurements, we performed an independent RF measurement based on a two-tone spectroscopy\cite{Krupko2018}.
This method  relies on measuring the dispersion relation of a resonator whose resonance frequency is set intentionally low, here $f_0 = \SI{750}{\mega\hertz}$. This allows to probe a large number of its harmonics.
To map the dispersion relation, a VNA is set to measure the transmission at a resonant frequency of the resonator $f_\mathrm{VNA}$ within the 4-\SI{8}{\giga\hertz} band of our measurement setup.
We then sweep a second tone at a frequency $f_\mathrm{MW}$ and whenever that second tone matches a harmonic of the resonator, at a frequency $f_\mathrm{n}$, the measured resonance at $f_\mathrm{VNA}$ is dispersively shifted by the cross-Kerr effect\cite{Suchoi2010} and the transmission readout by the VNA is modified.
By identifying all $f_\mathrm{n}$, the dispersion relation can be reconstructed.
In Fig.~\ref{fig:kinetic}(b)  we show the dispersion relation for a probe frequency $f_\mathrm{VNA} = \SI{5.22}{\giga\hertz}$,  the seventh harmonic of the resonator.
Since the angular wavenumber of each resonance is given by $k_\mathrm{n} = \pi  n/\ell$ where $\ell$ is the length of the $\lambda /2$ resonator and $n$ is the mode index, we can extract the kinetic inductance as follows:

\begin{equation}
     v_\mathrm{ph} = \dfrac{\omega_\mathrm{n}}{k_\mathrm{n}} = \dfrac{1}{\sqrt{ C_\ell ( L^\mathrm{m}_\ell + L^\mathrm{kin}_\ell)}},
\end{equation}
where $\omega_\mathrm{n} = 2\pi f_\mathrm{n}$ is the angular resonance frequency, $C_\mathrm{\ell}$ is the capacitance per unit length and $L^m_\ell$, $L^\mathrm{kin}_\ell$ are the geometric and the kinetic inductance per unit length respectively.
$L^\mathrm{m}_\ell$ and $C_\ell$ are purely geometrical quantities and can be estimated using a microwave simulation software like Sonnet ($L^\mathrm{m}_\ell = \SI{2.13e-7}{\henry\per\meter}$ and $C_\ell = \SI{2.82e-10}{\farad\per\meter}$) or conformal mapping calculations\cite{Simons2001} ($L^\mathrm{m}_\ell = \SI{2.13e-7}{\henry\per\meter}$ and $C_\ell = \SI{3.13e-10}{\farad\per\meter}$).
From this RF measurement and Sonnet simulations data we obtained a kinetic inductance value $L^\mathrm{kin}_\ell = \SI{3.84e-6}{\henry\per\meter}$ corresponding to $L_\mathrm{kin} = \SI{192(3)}{\pico\henry\per\sq}$, which is in excellent agreement with the DC measurement extraction.
The sheet inductance extracted is in line with the state-of-the-art values for NbN thin layers\cite{Annunziata2010, Adamyan2015, DeGraaf2018, Niepce2019}. Note that the kinetic inductance can be tuned by varying the substrate temperature during the sputter deposition. We find that by tuning the temperature from room temperature to \SI{275}{\degree C} the kinetic inductance changes from \SI{174}{\pico \henry \per \sq} to \SI{36}{\pico \henry \per \sq} as $T_\mathrm{c}$ evolves from \SIrange{5.6}{10.5}{K} (see the Supplementary Material). We stress that theses values are lower bounds of the kinetic inductance as it is estimated from the room temperature sheet resistance instead of the maximum $R_\Box$.

\begin{figure}
    \includegraphics{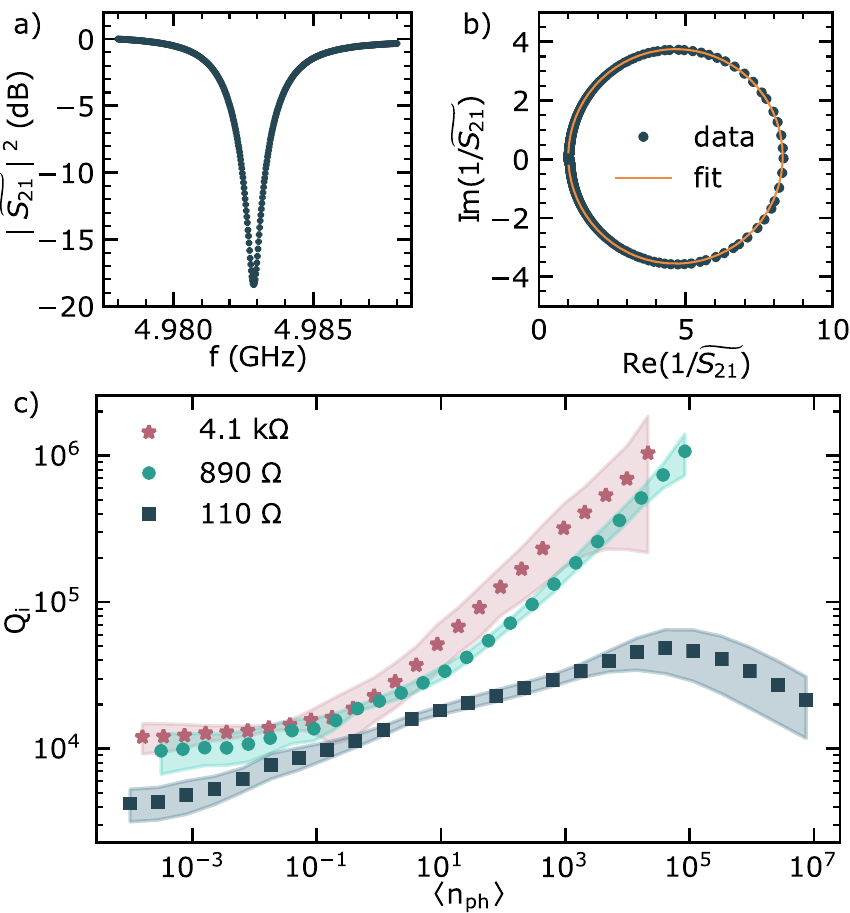}
    \caption{Power dependence of the resonator's internal quality factor.
    a) Typical normalized $\widetilde{ S_{21}}$ response of a resonator in the many photons regime.
    b) Parametric plot (dot) and fit (line) of $\operatorname{Im}(1/\widetilde{S_{21}})$ vs $\operatorname{Re}(1/\widetilde{S_{21}})$ (dots) of the same data as in a). 
    c) Power dependence of the internal quality factor for the resonators with different characteristic impedances. For each impedance, the internal quality factor is the mean value of four (\SI{4.1}{\kilo\ohm}) or five (\SI{110}{\ohm} and \SI{890}{\ohm}) resonators with the same impedance but different lengths (see the Supplementary Material for the data of each resonator). The shaded area corresponds to the standard deviation of the data.
    At low photon number, the internal quality factor saturates.
    At high photon number, $Q_\mathrm{i}$ for the \SI{110}{\ohm} resonator saturates at $\langle n_\mathrm{ph} \rangle = 10^5$ then decreases, which may be due to the high non-linearity of the material at high power.
    For the high-impedance resonators, the saturation of $Q_\mathrm{i}$ is not probed since it happens at very high power which makes the resonance unstable.}
    \label{fig:power}
\end{figure}

\setlength{\tabcolsep}{8pt}
\begin{table}[hbtp]
        \centering
        \caption{\label{tab:geometry} Characteristic impedance of the resonators with different geometries}
    \begin{tabular}{llll}
    \toprule[1.5pt]
        $Z_\mathrm{c}$     & \SI{110}{\ohm} & \SI{890}{\ohm} & \SI{4.1}{\kilo\ohm} \\ 
        \hline
        s  (\SI{}{\micro\meter}) & 50 & 2 & 0.2  \\ 
        w  (\SI{}{\micro\meter}) & 2  & 2 & 2  \\
    \bottomrule[1.5pt]
    \end{tabular}
\end{table}

From the NbN layer characterized previously (\SI{192}{\pico \henry \per \sq}) we have designed three sets of resonators with impedances of \SI{110}{\ohm}, \SI{890}{\ohm} and \SI{4.1}{k\ohm} by just varying the central conductor width $s$ from \SI{50}{\um} to \SI{200}{\nano\meter} while keeping the gap width $w= \SI{2}{\um}$, see Tab.~\ref{tab:geometry}.
We study the effect of the impedance and the input power on the internal quality factor of the resonators resonating in the 4-\SI{8}{GHz} band of our set-up.
A typical normalized transmission spectrum is shown in  Fig.~\ref{fig:power}(a).
The transmission spectrum is normalized by setting the background signal to \SI{0}{dB} and removing the electronic delay and phase shift of the measurement set-up.
Once normalized and close to resonance, the $\widetilde{S_\mathrm{21}}$ coefficient can be described by \cite{Megrant2012}

\begin{equation}
     \dfrac{1}{\widetilde{S_{21}}} = 1 + \dfrac{Q_\mathrm{i}}{Q_\mathrm{c}}e^{i\phi}\dfrac{1}{1 +i2Q_\mathrm{i}\delta x},
\end{equation}
where $Q_\mathrm{i}$, $Q_\mathrm{c}$ are the internal and coupling quality factor respectively, $\phi$ is the rotation in the $1/\widetilde{S_{21}}$ complex plane due to an impedance mismatches of the feedline of the resonators and $\delta x = (f-f_0)/f_0$ is the relative frequency to the resonance frequency $f_0$.
The fit is performed in the $1/\widetilde{S_{21}}$ complex plane to take into account the resonance response in magnitude and in phase simultaneously.
A typical result of such a fit is shown in Fig.~\ref{fig:power}(b).

From a circuit model (see Supplementary Material) we derive the average photon number inside the resonator as
\begin{equation}
    \langle n_\mathrm{ph}\rangle =  \frac{Q_\mathrm{c}}{\omega_0}\left(\dfrac{Q_\mathrm{i}}{Q_\mathrm{i}+Q_\mathrm{c}}\right)^2\dfrac{P_\mathrm{in}}{\hbar \omega_0},
\end{equation}
where $P_\mathrm{in}$ is the input power at the resonator and $\omega_0 = 2\pi f_0$.
Fig.~\ref{fig:power}(c) shows the average and the standard deviation of the internal quality factors of several resonators of a given impedance as a function of the average photon number. The individual data for the resonators with different impedances can be found in the Supplementary Material.
Before going into detail we precise that the \SI{110}{\ohm} resonators stayed in ambient atmosphere for a few months during the COVID-19 pandemic between its fabrication and the characterization, which may explain its different behaviour from the two other sets of resonators.
At low power we observe for the \SI{900}{\ohm} and \SI{4.5}{\kilo\ohm} resonators a clear saturation of the internal quality factor that may be explained by two-level system dynamics\cite{Burnett2017}.
At high photon number, $\langle n_\mathrm{ph}\rangle >10^5$, and for the same set of resonators, self-Kerr non-linearities lead to a strong asymmetry of the measured resonances rendering the analysis of the quality factors beyond the scope of our study.
We can only conclude that the internal quality factor saturation usually observed at such input powers\cite{Sharrow2018} was not visible up to the power shown here.
For the \SI{110}{\ohm} resonators, whom we suspect had a different aging evolution than the other sets of resonators, we do not observe a saturation in the single photon regime while observing a clear saturation at high power around $\sim 10^4$ photons.

\begin{figure}
\includegraphics{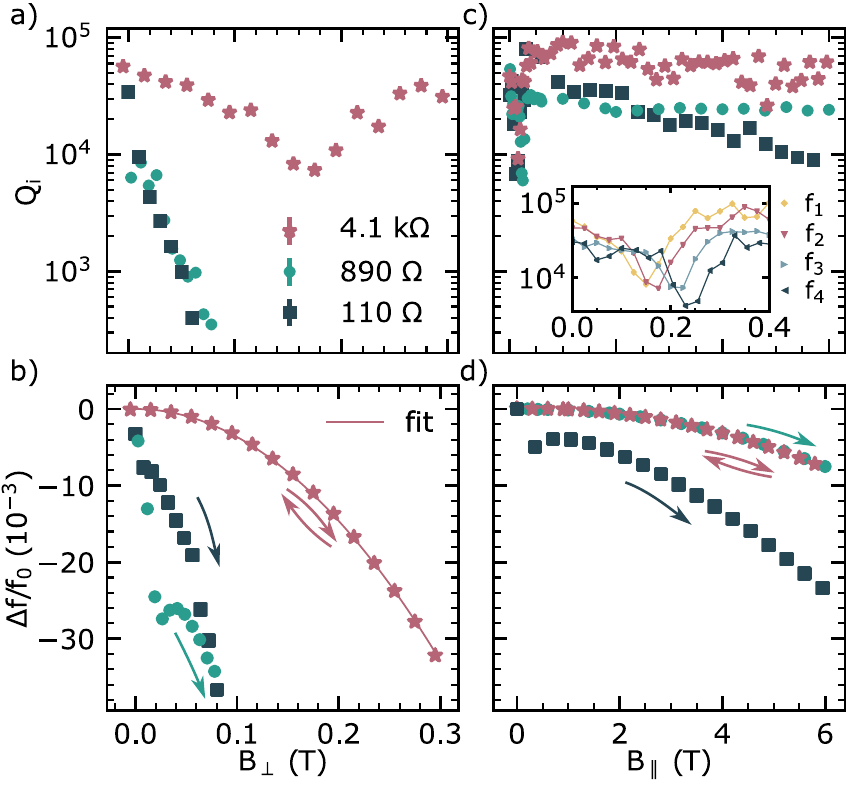}
\caption{\label{fig:field} Evolution of the resonators characteristics with a magnetic field.
a) $Q_\mathrm{i}$ as a function of $B_\perp$.
b) Normalized shift of the resonance frequencies with $B_\perp$ where the lower impedance resonators show abrupt jumps around zero field while the highest impedance resonator shows no abrupt jumps neither a hysteresis.
c) $Q_\mathrm{i}$ as a function of $B_\parallel$ for different impedances.
For the resonators at \SI{110}{\ohm} and \SI{4.1}{k\ohm}, the measurement is performed from 0 to \SI{-6}{\tesla} and for practical reasons it is plotted as positive values.
At low magnetic field, the dip in $Q_\mathrm{i}$ is due to coupling to magnetic impurities.
Inset: $Q_\mathrm{i}$ as a function of $B_\parallel$ around \SI{200}{\milli\tesla} for $Z_\mathrm{c} = \SI{4.1}{k\ohm}$ resonators resonating at different frequencies ($f_1 = \SI{3.8}{GHw}$, $f_2 = \SI{4.4}{\giga\hertz}$, $f_3 = \SI{5.6}{\giga\hertz}$, $f_4 = \SI{6.2}{\giga\hertz}$) all coupled to magnetic impurities of $g = 2$.
d) Normalized shift of the resonance frequencies with $B_\parallel$. Arrows in b) and d) indicate the sweep direction of the magnetic field.}
\end{figure}

We now turn to the behaviour of the resonators in a static magnetic field.
The internal quality factor and the relative frequency shift as a function of the applied magnetic field have been measured with an average of $\approx 100$ photons and the results can be seen in Fig.~\ref{fig:field}.
For an out-of-plane magnetic field, see Fig.~\ref{fig:field}(a) and (b), the internal quality factor drops to $10^2$ at \SI{100}{\milli\tesla} with an abrupt jump in resonance frequency around \SI{0}{\tesla} for the \SI{110}{\ohm} and \SI{890}{\ohm} resonators.
For the narrowest resonators (\SI{4.1}{k\ohm}) $Q_\mathrm{i}$ stays well above $10^4$ up to $B_\perp = \SI{300}{\milli\tesla}$ without any jump or hysteresis in the resonance frequency.
We note a dip in $Q_\mathrm{i}$ around $\sim \SI{150}{\milli\tesla}$, which can be associated with a coupling of the resonator to magnetic impurities.
The quadratic shift of the resonance is explained by the superconducting depairing under magnetic field and can be fitted following the expression\cite{Samkharadze2016} $\Delta f/f_0 = -(\pi/48)[De^2/(\hbar k_\mathrm{B}T_\mathrm{c})]B_\perp^2s^2$ with $D$ the electronic diffusion constant.
The extracted diffusion constant $D \approx \SI{0.58}{cm^2/s}$ is consistent with previous measurements\cite{Engel2013,Knehr2019} on NbN thin films.

For the in-plane magnetic field resilience, see Fig.~\ref{fig:field}(c) and (d), we find $Q_\mathrm{i} > 10^4$ for all resonators from \SI{500}{\milli\tesla} to \SI{6}{\tesla}.
Finally, both out-of-plane and in-plane magnetic field studies show that the highest impedance have also the highest magnetic field resilience. As the losses induced in a magnetic field are mainly attributed to the creation of magnetic-flux vortices in the superconducting film, a smaller width of the central conductor minimizes vortices creation and dynamics, thus suppressing the quality factor degradation. 
For the \SI{4.1}{\kilo\ohm} resonator for example, the central conductor width (\SI{200}{\nano\meter}) is shorter than the London penetration depth of NbN\cite{Kamlapure2010}. Therefore, vortices are induced only in the ground plane\cite{Kroll2019}, which explains its high $Q_\mathrm{i}$ in magnetic fields. The relative resonance shift in $B_\parallel$ in Fig.~\ref{fig:field}(d) shows that the \SI{110}{\ohm} resonances jump abruptly around \SI{0}{T}, which is a signature of unstable magnetic-flux vortices in the superconducting film, while the \SI{890}{\ohm} and the \SI{4.1}{\kilo\ohm} resonators show smooth shift of the resonance frequency and no hysteretic behaviour.
Thus, even without complex microwave engineering to minimize vortices dynamics, a nanowire CPW design already improves the magnetic resilience by several orders of magnitude for both in-plane and out-of-plane magnetic field.
In addition, we have verified that the excellent behaviour under a magnetic field of $B_\parallel$ = \SI{6}{T} with $Q_\mathrm{i}>10^4$ is preserved in the single photon regime.

Concerning the interaction between the resonator and magnetic impurities, the inset in Fig.~\ref{fig:field}(c) shows the internal quality factors of four \SI{4.1}{k\ohm} resonators with different resonance frequencies.
The observed dip in the internal quality factor is shifting to a higher magnetic field as the resonator frequency is increased as expected for the resonant condition  $g\mu_\mathrm{B} B = \hbar \omega_0$ with $g$ the Landé g-factor of the magnetic impurities. From all resonator measurements, we extract $g = 1.97 \pm 0.29$, which matches the g-factor of free electrons ($g = 2$).

In conclusion, we fabricate CPW resonators from a \SI{10}{\nano\meter} thick NbN film in a single e-beam lithography step with a kinetic inductance of \SI{192}{\pico\henry\per\sq} on silicon oxide.
The highest impedance reaches \SI{4.1}{\kilo\ohm}, which should induce zero-point voltage fluctuation one order of magnitude higher than for a \SI{50}{\ohm} resonator. The high kinetic inductance enables the fabrication of superinductor with relaxed geometry constraints compared to previous reports\cite{Niepce2019}.
We find, at zero field, an internal quality factor $Q_\mathrm{i} > 10^4$ in the single photon regime.
The narrow center conductor of the \SI{4.1}{k\ohm} resonator made it highly resilient to magnetic fields with $Q_\mathrm{i}>10^4$ in a \SI{300}{\milli\tesla} out-of-plane and in a \SI{6}{\tesla} in-plane magnetic field without any hysteresis in the resonance frequency.
Finally, the NbN HISMR is a compelling choice for cQED experiments operating at finite magnetic fields and involving quantum systems with small electric dipole moments.\\

\section*{Supplementary material}
See Supplementary Material for the table of NbN depositions performed at different substrate temperatures, the data of all resonators used for Fig.~\ref{fig:power}(c), and the derivation of the average photon number in a $\lambda/2$ resonator.

\begin{acknowledgments}
We acknowledge useful discussions with Romain Albert, Franck Balestro, Jérémie Viennot and Nicolas Roch. We also thank Michel Boujard for all mechanics parts to build the set-up. This work is supported by the ERC starting grant LONGPSIN (No.759388). S. Zihlmann acknowledges support by an Early Postdoc.Mobility fellowship (P2BSP2\_184387) from the Swiss National Science Foundation. G. Troncoso acknowledges the European Union’s Horizon 2020 research and innovation programme under the Marie Skłodowska-Curie grant agreement No.754303.
\end{acknowledgments}

\section*{Data availability}
The data that support the findings of this study are available from the corresponding authors upon reasonable request.

\bibliography{ref}
\end{document}


\title{Magnetic field resilient high kinetic inductance superconducting niobium nitride coplanar waveguide resonators: Supplementary material}

\author{Cécile Xinqing Yu$^1$, Simon Zihlmann$^1$, Gonzalo Troncoso Fernández-Bada$^1$, Jean-Luc Thomassin$^1$,\\
Frédéric Gustavo$^1$, \'{E}tienne Dumur$^1$ and Romain Maurand$^{1}$}
\date{%
$^1$Univ. Grenoble Alpes, CEA, Grenoble INP, IRIG, PHELIQS, F-38000 Grenoble, France\\}

\maketitle

\def\theequation{S\arabic{equation}}
\renewcommand{\thetable}{S\Roman{table}}
\renewcommand{\thefigure}{S\arabic{figure}}

\section{NbN films grown at different substrate temperature}

In this section, we summarize all the different NbN depositions we have made at different growth temperatures to investigate on its influence on the kinetic inductance of the film. All films are \SI{10}{\nano \meter} thick and when the heater is used, it is for $\approx\SI{16}{h}$. 

\setlength{\tabcolsep}{8pt}
\begin{table}[hbtp]
        \centering
        \caption{\label{tab:geometry} NbN film characterization for different sputtering temperatures.}
    \begin{tabular}{lllll}
    \toprule[1.5pt]
         Temperature (\SI{}{\celsius}) & $R_\Box$ (\SI{}{\ohm \per \sq}) & $T_\mathrm{c}$ (\SI{}{\kelvin}) & Kinetic inductance (\SI{}{\pico \henry \per \sq})\\ 
        \hline
        275 & 273 & 10.5 & \phantom{1}36   \\ 
        275 & 339 & \phantom{1}9.6 & \phantom{1}49   \\
        180 & 480 & \phantom{1}7.7 & \phantom{1}86 \\
        180  & 550  & \phantom{1}7.4 & 103  \\
        \phantom{1}22  & 670  & \phantom{1}5.9 & 162  \\
        \phantom{1}22  & 705  & \phantom{1}5.6 & 174  \\
    \bottomrule[1.5pt]
    \end{tabular}
\end{table}
Note that the kinetic inductance in Tab.~\ref{tab:geometry} is an estimate from the sheet resistances at room temperature from Eq.1 of the main paper. It is a lower bound of the real kinetic inductance as the resistance generally increases from \SI{300}{K} to the superconducting transition due to the disordered properties of NbN. 
For instance, the film sputtered at \SI{180}{\celsius} with $T_\mathrm{c} = \SI{7.4}{\kelvin}$ is the one used for the resonators characterization, and it sheet resistance before $T_\mathrm{c}$ is $R_\Box = \SI{1033}{\ohm \per \sq}$. 
The difference between films with the same substrate temperature  might be due to the variability of the sputter system and/or the accuracy of the film thickness.

\section{Power dependence of the resonators' internal quality factors}

\begin{figure}[h!]
\centering
\includegraphics{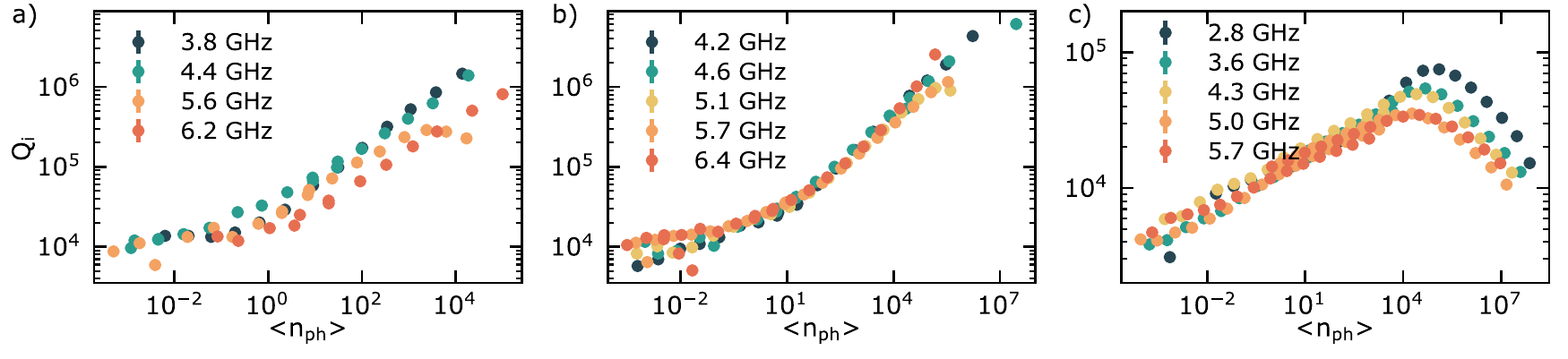}
\caption{Full dataset of the internal quality factors as a function of the average photon number in the resonators. a) $Q_\mathrm{i}$ as a function of $\langle n_\mathrm{ph}\rangle$ for the \SI{4.1}{\kilo\ohm} resonators with different fundamental resonance frequencies. b) $Q_\mathrm{i}$ as a function of $\langle n_\mathrm{ph}\rangle$ for the \SI{890}{\ohm} resonators. c) $Q_\mathrm{i}$ as a function of $\langle n_\mathrm{ph}\rangle$ for the \SI{110}{\ohm} resonators.}
\label{fig:Qi}
\end{figure}

In Fig.~\ref{fig:Qi}, we show the full dataset of the internal quality factor as a function of the input power for all resonators individually.

\section{Average number of photons in a $\lambda/2$ resonator}

\begin{figure}[h!]
\centering
\includegraphics{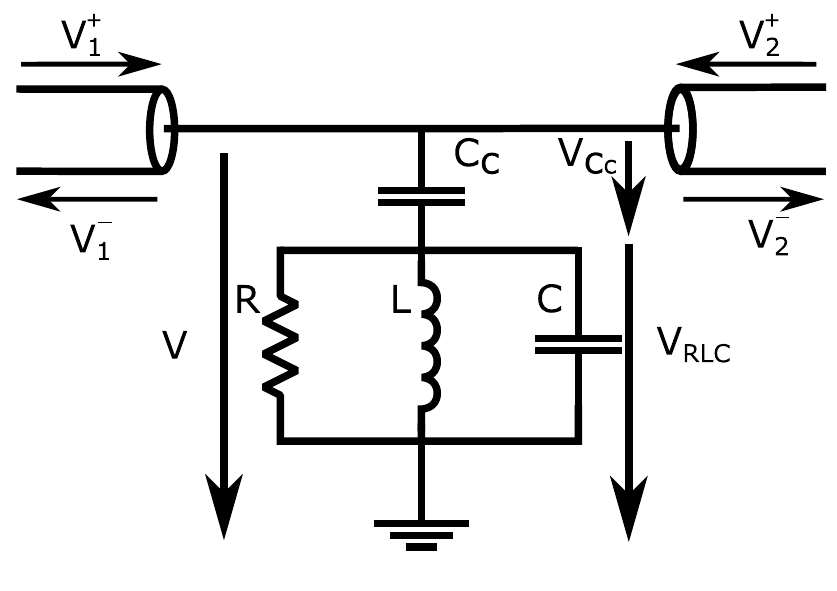}
\caption{Electrical circuit model of a parallel resonator, as a RLC circuit, coupled to a feedline via a coupling capacitor $C_\mathrm{c}$. }
\label{fig:model}
\end{figure}

In this section we will derive the internal average photon number of a resonator driven by an external drive as stated in Eq.~4 of the main paper.
As shown in Ref~\cite{pozar} section 6.2, a low loss $\lambda/2$ resonator may be described as a parallel RLC resonator at frequencies close to its resonance frequency.
From there we also assume a purely capacitive coupling between the resonator and its feedline, an assumption justified by the electro-magnetic field distribution along the resonator and the position of the resonator in respect to the feedline.
The resulting circuit model is shown in Fig.~\ref{fig:model}.

From Ref.~\cite{pozar}, Eq.~6.14.b, the average electric energy stored in the capacitor $C$ is
\begin{equation}
W_\mathrm{e} = \frac{1}{4} C |V_\mathrm{RLC}|^2.
\end{equation}

The aim of this section is to derive an expression of $W_\mathrm{e}$ as a function of the resonator resonance frequency $\omega_\mathrm{r}$, its internal quality factor $Q_\mathrm{i}$ and its coupling quality factor $Q_\mathrm{c}$, all of them being accessible through fitting of the resonance spectrum.

\subsection{Derivation of $V_\mathrm{RLC}$}

From Ohm's law we have
\begin{equation}
V_\mathrm{RLC} =Z_\mathrm{RLC} I.
\end{equation}
Using the Kirchhoff's second law, we can calculate the applied current as
\begin{align}
V & = V_{C_\mathrm{c}} + V_\mathrm{RLC}, \\
V & = I ( Z_{C_\mathrm{c}} +Z_\mathrm{RLC}),\\
I & = \frac{V}{ Z_{C_\mathrm{c}} +Z_\mathrm{RLC}}.
\end{align}
From which the voltage reads
\begin{equation}
V_\mathrm{RLC} = \frac{Z_\mathrm{RLC}}{ Z_{C_\mathrm{c}} +Z_\mathrm{RLC}}V.
\label{eq:Vrlc}
\end{equation}
From Ref.~\cite{pozar} Eq.~4.41, we know that: 
\begin{align}
V_2^- & = S_{21} V_1^+ ,\\
V & = S_{21} V_1^+.
\label{eq:S21}
\end{align}
By injecting Eq.~\ref{eq:S21} to Eq.~\ref{eq:Vrlc}, we obtain 
\begin{equation}
V_\mathrm{RLC} =  \frac{Z_\mathrm{RLC}}{ Z_{C_\mathrm{c}} +Z_\mathrm{RLC}}S_{21}V_1^+.
\end{equation}
Finally, the average energy can be written as
\begin{equation}
W_\mathrm{e} = \frac{1}{4} C \left| \frac{Z_\mathrm{RLC}}{ Z_{C_\mathrm{c}} +Z_\mathrm{RLC}} \right|^2 | S_{21}|^2 |V_1^+|^2.
\label{eq:We1}
\end{equation}
The input power of the system is, from Ref.~\cite{pozar} Eq.~6.13,
\begin{equation}
P_\mathrm{in} = \frac{1}{2}\frac{|V_1^+|^2}{Z_0}.
\label{eq:Pin}
\end{equation}
By using Eq.~\ref{eq:Pin} in Eq.~\ref{eq:We1}, we obtain 
\begin{equation}
W_\mathrm{e} = \frac{Z_0}{2} C \left| \frac{Z_\mathrm{RLC}}{ Z_{C_\mathrm{c}} +Z_\mathrm{RLC}} \right|^2 | S_{21}|^2 P_\mathrm{in}.
\label{eq:We2}
\end{equation}

\subsection{Derivation of resonator impedance}

The impedance of an open-circuited $\lambda/2$ resonator is given by Ref.~\cite{pozar} Eq.~6.33 and we will express it through its internal quality factor, see Ref.~\cite{pozar}, Eq.~6.35, as
\begin{equation}
Z_\mathrm{RLC} = \frac{Z_\mathrm{r}}{\frac{\pi}{2} \frac{1}{Q_\mathrm{i}} + i\pi \frac{\Delta\omega_\mathrm{r}}{\omega_\mathrm{r}}},
\end{equation}
with $Z_\mathrm{r}$, the resonator characteristic impedance and $\Delta\omega_\mathrm{r} = \omega - \omega_\mathrm{r}$, the drive frequency relative to the resonance frequency.
We rewrite the previous equation by separating the real and imaginary part
\begin{equation}
Z_\mathrm{RLC} = Z_\mathrm{r} \frac{2Q_\mathrm{i}}{\pi} \frac{1 - i2Q_\mathrm{i}\Delta\omega_\mathrm{r}/\omega_\mathrm{r}}{1+4Q_\mathrm{i}^2\left(\frac{\Delta\omega_\mathrm{r}}{\omega_\mathrm{r}}\right)^2}.
\label{eq:rlc}
\end{equation}

\subsection{The resonance shift}

While the bare resonator resonates at a frequency $\omega_\mathrm{r}$ the resonator coupled to a feedline does so at a frequency $\omega_0$, lower than $\omega_\mathrm{r}$.
In this section we derive $\Delta\omega_\mathrm{0}/\omega_\mathrm{0}$ the drive frequency relative to the resonance of the resonator coupled to the feedline.
The total impedance is 
\begin{equation}
Z_\mathrm{tot} = Z_\mathrm{r} \frac{2Q_\mathrm{i}}{\pi} \frac{1 - i2Q_\mathrm{i}\Delta\omega_\mathrm{r}/\omega_\mathrm{r}}{1+4Q_\mathrm{i}^2\left(\frac{\Delta\omega_\mathrm{r}}{\omega_\mathrm{r}}\right)^2}  - \frac{i}{\omega C_\mathrm{c}},
\end{equation}

\begin{equation}
Z_\mathrm{tot} = Z_\mathrm{r} \frac{\frac{2Q_\mathrm{i}}{\pi} - i \left( \frac{4Q_\mathrm{i}^2}{\pi}\frac{\Delta\omega_\mathrm{r}}{\omega_\mathrm{r}} + \frac{1}{\omega Z_\mathrm{r} C_\mathrm{c}} \left[1 + 4Q_\mathrm{i}^2 \left(\frac{\Delta\omega_\mathrm{r}}{\omega_\mathrm{r}}\right)^2\right]\right)}{1 + 4Q_\mathrm{i}^2 \left(\frac{\Delta\omega_\mathrm{r}}{\omega_\mathrm{r}}\right)^2 }.
\label{eq:ztot2}
\end{equation}
At resonance, the imaginary part of the impedance is equal to zero which leads to
\begin{equation}
\frac{4Q_\mathrm{i}^2}{\omega_0 Z_\mathrm{r} C_\mathrm{c}}\left(\frac{\Delta\omega_\mathrm{r}}{\omega_\mathrm{r}}\right)^2 + \frac{4Q_\mathrm{i}^2}{\pi} \frac{\Delta\omega_\mathrm{r}}{\omega_\mathrm{r}} + \frac{1}{\omega_0 Z_\mathrm{r} C_\mathrm{c}} = 0 .
\end{equation}
Assuming\footnote{By substituting $C_\mathrm{c}$ using Eq.~\ref{eq:qc}, this assumption is equivalent to $Q_\mathrm{i}^2/Q_\mathrm{c}^2 \gg\pi^2 Z_0/Z_\mathrm{r}$ which is often the case in cQED experiment with under-coupled resonators.} that $Q_\mathrm{i}^2/\pi^2 \gg 1/\omega^2Z_\mathrm{r}^2C_\mathrm{c}^2$ , the roots of the equation are
\begin{equation}
\left(\frac{\Delta \omega_\mathrm{r}}{\omega_\mathrm{r}}\right)_\pm = \frac{\omega_0 Z_\mathrm{r} C_\mathrm{c}}{2Q_\mathrm{i}}\left( - \frac{Q_\mathrm{i}}{\pi} \pm \frac{Q_\mathrm{i}}{\pi}\right).
\end{equation}
The only physical solution is the one lowering the resonance frequency which is consistent with adding the coupling capacitor in series with the resonator.
From this we define the drive frequency relative to the resonance of the coupled resonator as
\begin{equation}
\frac{\Delta \omega_0}{\omega_0} = \frac{\Delta \omega_\mathrm{r}}{\omega_\mathrm{r}} + \frac{\omega_0 Z_\mathrm{r} C_\mathrm{c}}{\pi}.
\label{eq:dr}
\end{equation}

\subsection{Coupling quality factor}

At resonance, the total impedance is, from Eq.~\ref{eq:ztot2},
\begin{equation}
Z_\mathrm{tot} = Z_\mathrm{r}\frac{2Q_\mathrm{i}}{\pi}\frac{1}{1+4Q_\mathrm{i}^2\left(\frac{\Delta \omega_\mathrm{r}}{\omega_\mathrm{r}}\right)^2}.
\end{equation}
To which we substitute Eq.~\ref{eq:dr}, leading to
\begin{align}
Z_\mathrm{tot} & = Z_\mathrm{r}\frac{2Q_\mathrm{i}}{\pi}\frac{\pi^2}{\pi^2+4Q_\mathrm{i}^2\omega^2 Z_\mathrm{r}^2 C_\mathrm{c}^2}.
\end{align}
Assuming $\pi \ll 4Q_\mathrm{i}^2\omega^2Z_\mathrm{r}^2C_\mathrm{c}^2$ :
\begin{align}
Z_\mathrm{tot} & \approx \frac{\pi}{2Q_\mathrm{i}\omega^2 Z_\mathrm{r} C_\mathrm{c}^2},
\label{eq:ztot}
\end{align}
 This assumption require a quality factor $\gg 10^3$ for giga-hertz \SI{50}{\ohm} resonators and a lower quality factor for higher impedances. In practice, for superconducting resonators used in typical cQED experiments this assumption is often verified. \\
From Ref.~\cite{pozar}, Table~4.2, we write the transmission coefficient as
\begin{align}
S_{21} & = \frac{2}{2+\frac{Z_0}{Z_\mathrm{tot}}}
\label{eq:S21_Z}\\
S_{21} & = \frac{2\pi}{2\pi+2Q_\mathrm{i}\omega^2 Z_\mathrm{r} C_\mathrm{c}^2 Z_0}.
\end{align}
At resonance, the transmission reaches a minimum
\begin{equation}
    S_{21}^\text{res} = \frac{Q_\mathrm{c}}{Q_\mathrm{c} + Q_\mathrm{i}},
    \label{eq:s21min}
\end{equation}
which allows us to derive the coupling quality factor by identification as 
\begin{equation}
Q_\mathrm{c} = \frac{\pi}{\omega^2 Z_\mathrm{r} C_\mathrm{c}^2 Z_0}.
\label{eq:qc}
\end{equation}

\subsection{Average number of photons in a $\lambda/2$ resonator}

Substituting Eq.~\ref{eq:qc} in Eq.~\ref{eq:dr} and in Eq.~\ref{eq:rlc}, we can write the resonator impedance at resonance frequency as
\begin{align}
   Z_\mathrm{RLC} & = Z_\mathrm{r}\frac{2Q_\mathrm{i}}{\pi}\frac{1}{1-2Q_\mathrm{i}\sqrt{\frac{Z_\mathrm{r}}{\pi Z_0 Q_\mathrm{c}}}}.
\end{align}
Assuming $1 \ll 2 Q_\mathrm{i}\sqrt{\frac{Z_\mathrm{r}}{\pi Z_0 Q_\mathrm{c}}}$ which is often true in cQED experiments where the resonator is under-coupled with $Q_\mathrm{i} \gg Q_\mathrm{c}$ :
\begin{equation}
    Z_\mathrm{RLC} \approx  -\sqrt{\frac{Z_\mathrm{r} Z_0 Q_\mathrm{c}}{\pi}}.
    \label{eq:zrlc2}
\end{equation}
Rearranging Eq.~\ref{eq:S21_Z} we obtain
\begin{equation}
\frac{1}{Z_{C_\mathrm{c}}+Z_\mathrm{RLC}} = \frac{2}{Z_0} \left( \frac{1}{S_{21}} - 1 \right).
\label{eq:Zs}
\end{equation}
The capacitance of the $\lambda/2$ resonator is, see Ref.~\cite{pozar} Eq.6.34.b,
\begin{equation}
C = \frac{\pi}{2 Z_\mathrm{r}\omega_\mathrm{r}}.
\label{eq:c}
\end{equation}
Substituting Eq.~\ref{eq:s21min}, Eq.~\ref{eq:zrlc2}, Eq.~\ref{eq:Zs}, Eq.~\ref{eq:c},  and  in Eq.~\ref{eq:We2} we finally obtain
\begin{equation}
W_\mathrm{e} = \frac{Q_\mathrm{c}}{\omega_0} \left( \frac{Q_\mathrm{i}}{Q_\mathrm{i}+Q_\mathrm{c}}\right)^2 P_\mathrm{in}.
\label{eq:We3}
\end{equation}
From which we deduce the average internal photon number in the resonator as
\begin{equation}
\langle n_\mathrm{photons}\rangle = \frac{Q_\mathrm{c}}{\omega_0} \left( \frac{Q_\mathrm{i}}{Q_\mathrm{i}+Q_\mathrm{c}}\right)^2 \frac{P_\mathrm{in}}{\hbar\omega_0}.
\end{equation}

\bibliographystyle{unsrt}
\bibliography{ref_sup.bib}